\documentclass[12pt]{article}
\usepackage{setspace}
\usepackage{epsfig}
\usepackage{amssymb}
\usepackage{graphicx}
\def\be{\begin{equation}}
\def\ee{\end{equation}}
\def\ba{\begin{array}}
\def\ea{\end{array}}
\def\beqn{\begin{eqnarray}}
\def\eeqn{\end{eqnarray}}
\def\nonum{\nonumber}
\def\bt{\begin{tabular}}
\def\et{\end{tabular}}
\def\bc{\begin{center}}
\def\ec{\end{center}}

\begin{document}

\title{Textures and Lepton Mass Matrices}
\author{Priyanka Fakay$^*$, Samandeep Sharma $^{**}$\\\textit{$^*$Department of Physics, D.A.V. College, Sector-10,
 Chandigarh-160011}\\
\textit{ $^{**}$ Department of Physics, GGDSD College, Sector-32
C, Chandigarh-160030}\\ \textit{priyanka.fakay@gmail.com}}

\maketitle \abstract{In view of the precise measurement of the
leptonic mixing angle $\theta_{13}$, implications of the latest
mixing data have been investigated for lepton mass matrices
pertaining to Dirac neutrinos.These texture specific lepton mass
matrices have been examined for their compatibility with the
latest data in the cases of normal hierarchy, inverted hierarchy
and degenerate scenario of neutrino masses.}\vskip 0.75cm

The precise measurement of the neutrino mixing angle $
s_{13}$\cite{theta13} has added another dimension  to our
knowledge of neutrino oscillation phenomenology. Interestingly,
this unexpectedly large value of $ s_{13}$ almost near the Cabibbo
angle, seem to have
 important implications for flavor physics. This value which
is neither like the other two mixing angles nor canonical has made lepton mixing pattern more
complicated compared to expectations from hitherto believed  underlying symmetries of the mixing matrix.
The observation of non zero $ s_{13}$ value,one the one hand, restores the
parallelism between quark mixing and lepton mixing, while on the other hand
signifies the difference between the uniformly large mixing angles for leptons compared with  the
corresponding quark mixing angles.

    Ever since the measurement of $ s_{13}$, intense amount of
    activity has taken place in understanding phenomenology of the pattern of neutrino
    masses and mixings. Further, it becomes desirable to understand
    quark mixing and lepton mixing from similar perspectives,
    however, while following unified approach, one has to keep in
    mind the difference in the mixing patterns of the quarks and
    the leptons. In this context, one should note that unlike the case of  quark mixings
which show a hierarchical structure, the pattern of neutrino
mixings do not show any explicitly hierarchy. Further, at present
we have no clue about the hierarchy of neutrino masses which may
be normal/inverted  or may even be degenerate. Furthermore,
exploring the relationship between neutrino mixing angles and the
lepton mass matrices gets further complicated
 by the fact that at present it is not clear whether neutrinos are  Dirac or Majorana particles.

    In the absence of any viable theory for flavor
physics, one usually resorts to phenomenological models such as
texture specific mass matrices which have received a good deal of
attention in the literature, for details in this regard we refer
the reader to a recent review article . Texture specific mass
matrices were introduced implicitly by Weinberg\cite{fri5} and
explicitly by Fritzsch\cite{frzans}. In particular, Fritzsch-like
texture specific mass matrices seem to be very helpful in
understanding the pattern of quark mixings and CP violation.For
details we refer the readers to \cite{singreview} In view of the
significant difference between the quark mixing pattern and the
lepton mixing pattern, as well as in the absence of data regarding
mass pattern of neutrinos one may have to resort to detailed and
case by case analyses for all the neutrino mass hierarchies as
well as for both Majorana and Dirac neutrinos, keeping in mind
that Dirac neutrinos have not yet been ruled out by experiment .
To this end it should be noted that several
attempts\cite{gulsheendirac} have been made in exploring the
possibility  of Dirac neutrinos having small masses as well as
their compatibility with the supersymmetric GUTs.

    In the case of Majorana neutrinos after the recent measurement
    of $ s_{13}$, texture 6 and 5 zero mass matrices have been
    examined in detail\cite{priyanka}.  However similar extensive attempts for
    Dirac neutrinos have not yet been carried out. Keeping in mind
    the parallelism between quark and lepton mixing phenomena and
    noting that the Dirac neutrinos have not yet been ruled out,
    it becomes desirable
    to study texture specific Dirac neutrino mass matrices. In this
    context, it may be added that the original texture 6 zero Fritzsch-like  mass
    matrices have been ruled out in the case of quarks, similarly  a
    closer look at some of these attempts  indicate the same for texture 6 zero Dirac
    mass matrices\cite{ptep}.

        In the case of quarks it has been shown that texture 5 zero mass matrices
        have been largely ruled out however, a detailed analysis has
        not been carried out for the  case of Dirac neutrinos.
        Therefore it becomes  desirable to carry out detailed
        analyses of texture 5 zero Dirac neutrino mass matrices, this is
        particularly important in view of recent refinements and measurement of
        angle $ s_{13}$. The plan of present work
                   is as follows. Firstly to make the document
                   self contained we discuss the texture specific Dirac neutrino
                   mass matrices   in Section I. In Section II we discuss
                   the  inputs used for analyses. In Section III we present the complete
                   analyses of texture 5 zero Dirac mass matrices
                   for normal hierarchy, inverted hierarchy and
                   degenerate scenario. Lastly in Section IV we
                   summarize our conclusions.

    \section{Texture 5 zero Dirac neutrino mass matrices}
    To define the various texture specific cases considered here, we
begin with the modified Fritzsch-like mass matrices, for example,
 \be
 M_{l}=\left( \ba{ccc}
0 & A _{l} & 0      \\ A_{l}^{*} & D_{l} &  B_{l}     \\
 0 &     B_{l}^{*}  &  C_{l} \ea \right), \qquad
M_{\nu D}=\left( \ba{ccc} 0 &A _{\nu} & 0      \\ A_{\nu}^{*} &
D_{\nu} &  B_{\nu}     \\
 0 &     B_{\nu}^{*}  &  C_{\nu} \ea \right),
 \label{frzmm}
 \ee
$M_{l}$ and $M_{\nu D}$ respectively corresponding to Dirac-like
charged lepton and neutrino mass matrices. Both the matrices are
texture 2 zero type with $A_{l(\nu)}
=|A_{l(\nu)}|e^{i\alpha_{l(\nu)}}$
 and $B_{l(\nu)} = |B_{l(\nu)}|e^{i\beta_{l(\nu)}}$. The two possible cases of texture 5 zero
matrices can be obtained by  taking either $D_l=0$ and
$D_{\nu}\neq 0$ or $D_{\nu}=0$ and $D_l \neq 0$, referred to as
texture 5 zero $D_l=0$ case pertaining to $M_l$ texture 3 zero
type and $M_{\nu D}$ texture 2 zero type and texture 5 zero
$D_{\nu}=0$ case pertaining to $M_l$ texture 2 zero type and
$M_{\nu D}$ texture 3 zero type. The formalism connecting the mass
matrix to the neutrino mixing matrix involves diagonalization of
the mass matrices $M_l$ and $M_{\nu D}$ and the details in this
regard can be looked up in \cite{singreview}, however to
facilitate discussion of results we briefly present some of the
essentials in this regard. To facilitate diagonalization, the mass
matrix $M_k$, where $k=l, \nu D$ can be expressed as \be M_k= Q_k
M_k^r P_k \, \label{4mk} \ee or \be M_k^r= Q_k^{\dagger} M_k
P_k^{\dagger}\,, \label{mkr} \ee where $M_k^r$ is a real symmetric
matrix with real eigenvalues and $Q_k$ and $P_k$ are diagonal
phase matrices. The real matrix $M_k^r$ is diagonalized by the
orthogonal transformation $O_k$, for example, \be M_k^{diag}=
{O_k}^T M_k^r O_k \,, \label{mkdiag} \ee which on using equation
(\ref{mkr}) can be rewritten as \be M_k^{diag}= {O_k}^T
Q_k^{\dagger} M_k P_k^{\dagger} O_k \,. \label{mkdiag2} \ee

To understand the relationship between diagonalizing
transformations for different hierarchies of neutrino masses as
well as their relationship with the charged lepton case, we
reproduce the general diagonalizing transformation $O_k$. The
elements of $O_k$ can figure with different phase possibilities,
however these possibilities are related to each other through the
phase matrices. For the present work, we have chosen the
possibility, \be O_k= \left( \ba{ccc} ~~O_k(11)& ~~O_k(12)&
~O_k(13)
\\
 ~~O_k(21)& -O_k(22)& ~O_k(23)\\
     -O_k(31) & ~~O_k(32) & ~O_k(33) \ea \right), \ee
where,  \beqn O_k(11) & = & {\sqrt \frac{m_{2} m_{3}
(m_{3}-m_{2}-D_k)}
     {(m_{1}-m_{2}+m_{3}-D_k)
(m_{3}-m_{1})(m_{1}+m_{2})} } \nonum  \\ O_k(12) & = & {\sqrt
\frac{m_{1} m_{3}
 (m_{1}+m_{3}-D_k)}
   {(m_{1}-m_{2}+m_{3}-D_k)
 (m_{2}+m_{3})(m_{2}+m_{1})} }
\nonum   \\O_k(13) & = & {\sqrt \frac{m_{1} m_{2}
 (m_{2}-m_{1}+D_k)}
    {(m_{1}-m_{2}+m_{3}-D_k)
(m_{3}+m_{2})(m_{3}-m_{1})} } \nonum   \\ O_k(21) & = & {\sqrt
\frac{m_{1}
 (m_{3}-m_{2}-D_k)}
  {(m_{3}-m_{1})(m_{1}+m_{2})} }
\nonum  \\O_k(22) & = & {\sqrt \frac{m_{2} (m_{3}+m_{1}-D_k)}
  {(m_{2}+m_{3})(m_{2}+m_{1})} }
 \nonum    \\
O_k(23) & = & \sqrt{\frac{m_3(m_{2}-m_{1}+D_k)}
 {(m_{2}+m_{3})(m_{3}-m_{1})} }
\nonum   \\O_k(31) & = &
 \sqrt{\frac{m_{1} (m_{2}-m_{1}+D_k)
    (m_{1}+m_{3}-D_k)}
{(m_{1}-m_{2}+m_{3}-D_k)(m_{1}+m_{2})(m_{3}-m_{1})}} \nonum
\\O_k(32) & = & {\sqrt \frac{m_{2}(D_k-m_{1}+m_{2})
(m_{3}-m_{2}-D_k)}{(m_{1}-m_{2}+m_{3}-D_k)
 (m_{2}+m_{3})(m_{2}+m_{1})} }
 \nonum  \\
O_k(33) & = & {\sqrt \frac{m_{3}(m_{3}-m_{2}-D_k)
(m_{1}+m_{3}-D_k)}{(m_{1}-m_{2}+m_{3}-D_k)
 (m_{3}-m_{1})(m_{3}+m_{2})}} \label{4diageq} \,,
 \eeqn  $m_1$, $-m_2$,
$m_3$ being the eigenvalues of $M_k$.

In the case of charged leptons, because of the hierarchy $m_e \ll
m_{\mu} \ll m_{\tau}$, the mass eigenstates can be approximated
respectively to the flavor eigenstates. Using the approximation,
$m_{l1} \simeq m_e$, $m_{l2} \simeq m_{\mu}$ and $m_{l3} \simeq
m_{\tau}$, the first element of the matrix $O_l$ can be obtained
from the corresponding element of equation (\ref{4diageq}) by
replacing $m_1$, $-m_2$, $m_3$ with $m_e$, $-m_{\mu}$, $m_{\tau}$,
for example,
  \be  O_l(11) = {\sqrt
\frac{m_{\mu} m_{\tau} (m_{\tau}-m_{\mu}-D_l)}
     {(m_{e}-m_{\mu}+m_{\tau}-D_l)
(m_{\tau}-m_{e})(m_{e}+m_{\mu})} } ~. \ee

In the case of neutrinos, for normal hierarchy of neutrino masses
defined as $m_{\nu_1}<m_{\nu_2}\ll m_{\nu_3}$ as well as for the
corresponding degenerate case given by $m_{\nu_1} \lesssim
m_{\nu_2} \sim m_{\nu_3}$ equation (\ref{4diageq}) can also be
used to obtain the first element of diagonalizing transformation
for Dirac neutrinos. This element can be obtained from the
corresponding element of equation (\ref{4diageq}) by replacing
$m_1$, $-m_2$, $m_3$ with $m_{\nu 1}$, $-m_{\nu 2}$, $m_{\nu 3}$
and is given by
\be O_{\nu D}(11)  =  {\sqrt \frac{m_{\nu_2} m_{\nu 3} (m_{\nu 3}-m_{\nu 2}-D_{\nu})}
     {(m_{\nu 1}-m_{\nu 2}+m_{\nu 3}-D_{\nu})
(m_{\nu 3}-m_{\nu 1})(m_{\nu 1}+m_{\nu 2})} }, \ee where $m_{\nu_1}$, $m_{\nu_2}$ and $m_{\nu_3}$ are neutrino
masses.

In the same manner, one can obtain the elements of diagonalizing
transformation for the inverted hierarchy case defined as
$m_{\nu_3} \ll m_{\nu_1} < m_{\nu_2}$ as well as for the
corresponding degenerate case given by $m_{\nu_3} \sim m_{\nu_1}
\lesssim m_{\nu_2}$. The corresponding first element, obtained by
replacing $m_1$, $-m_2$, $m_3$ with $m_{\nu 1}$, $-m_{\nu 2}$,
$-m_{\nu 3}$ in equation (\ref{4diageq}) is given by
  \be O_{\nu D}(11)  =  {\sqrt \frac{m_{\nu_2} m_{\nu 3} (m_{\nu 3}+m_{\nu 2}+D_{\nu})}
     {(-m_{\nu 1}+m_{\nu 2}+m_{\nu 3}+D_{\nu})
(m_{\nu 3}+m_{\nu 1})(m_{\nu 1}+m_{\nu 2})} }. \ee The other elements of diagonalizing transformations in the
case of neutrinos as well as charged leptons can similarly be
found.

After the elements of diagonalizing transformations $O_l$ and
$O_{\nu D}$ are known, the Pontecorvo-Maki-Nakagawa-Sakata (PMNS)
matrix \cite{pmns} can be obtained through the relation
\be
 U = O_l^{\dagger} Q_l P_{\nu D} O_{\nu D} \,, \label{4mixreal} \ee
where $Q_l P_{\nu D}$, without loss of generality, can be taken as
$(e^{i\phi_1},\,1,\,e^{i\phi_2})$, $\phi_1$ and $\phi_2$ being
related to the phases of mass matrices and can be treated as free
parameters.

\section{Inputs used in the analysis}

The
present work uses results from the latest global three neutrino
oscillation analysis carried out by Fogli {\it et al}.
\cite{f2012}. At 1$ \sigma $ C.L. the allowed ranges of the
various input parameters are \be
 \Delta {\it m}_{21}^{2} = (7.32-7.80)\times
 10^{-5}~\rm{eV}^{2},~~~~
 \Delta {\it m}_{23}^{2} = (2.33-2.49)\times 10^{-3}~ \rm{eV}^{2},
 \label{deltamass1}\ee
\be
s^2 _{12}  =  (0.29-0.33),~~~
 s^2_{23}  =  (0.37-0.41),~~~
s^2 _{13} = (0.021-0.026), \label{angles1} \ee where $ \Delta {\it
m}_{ij}^{2}$ 's correspond to the solar and atmospheric neutrino
mass square differences and $ s_{ij}$ corresponds to the sine of
the mixing angle $ij$ where $i,j=1,2,3$. At 3$ \sigma $ C.L. the
allowed ranges are given as \be
 \Delta {\it m}_{21}^{2} = (6.99-8.18)\times
 10^{-5}~\rm{eV}^{2},~~~~
 \Delta {\it m}_{23}^{2} = (2.19-2.62)\times 10^{-3}~ \rm{eV}^{2},
 \label{deltamass2}\ee
\be
s^2 _{12}  =  (0.26-0.36),~~~
 s^2_{23}  =  (0.33-0.64),~~~
s^2 _{13} = (0.017-0.031). \label{angles2} \ee Based upon the
information regarding neutrino masses and mixing parameters,
Garcia {\it et al.} \cite{garcia2} have constructed the PMNS
matrix taking into account the neutrino oscillation data. For
example, the magnitudes of the elements of the PMNS matrix at
3$\sigma $ C.L. given by  Garcia {\it et al.} \cite{garcia2} are
\be {\rm V_{PMNS}} = \left(\ba{ccc}
  0.759~-~0.846 & 0.513~ -~ 0.585 & 0.126~-~0.178\\
 0.205 ~-~ 0.543  & 0.416 ~-~ 0.730  & 0.579~ -~0.808\\
 0.215 ~-~ 0.548 & 0.409 ~-~ 0.725 & 0.567~-~0.800
 \ea \right). \label{gar} \ee
It may be noted that
lightest neutrino mass corresponds to $m_{\nu_1}$ for the normal
hierarchy case and to $m_{\nu_3}$ for the inverted hierarchy case.
The lightest neutrino
mass, the phases $\phi_1$, $\phi_2$ and $D_{l, \nu}$ have been
considered as free parameters,in the normal hierarchy case the other two masses are
constrained by $\Delta m_{12}^2 = m_{\nu_2}^2 - m_{\nu_1}^2 $ and
$\Delta m_{23}^2 = m_{\nu_3}^2 - m_{\nu_2}^2 $  and by $\Delta m_{23}^2 = m_{\nu_2}^2 -
m_{\nu_3}^2$ in the inverted hierarchy case. The explored range for
$m_{\nu_1}$ is taken to be $10^{-8}\,\rm{eV}-10^{-1}\,\rm{eV}$.
 In the absence of any constraint i.e., in the absence of CP violation in the leptonic sector, phases, $\phi_1$
and $\phi_2$ have been given full variation from 0 to $2\pi$.
Although $D_{l, \nu}$ are free parameters, however, they have been
constrained such that diagonalizing transformations, $O_l$ and
$O_{\nu}$, always remain real, implying $D_{l}< m_{l_3} - m_{l_2}$
whereas $D_{\nu} < m_{\nu_3} - m_{\nu_2}$ for normal hierarchy and
$D_{\nu} < m_{\nu_1} - m_{\nu_3}$ for inverted hierarchy.

\section{Results and discussion}
\subsection{Inverted hierarchy of neutrino masses}
\begin{figure}[hbt]
\begin{minipage}{0.45\linewidth}   \centering
\includegraphics[width=2.in,angle=-90]{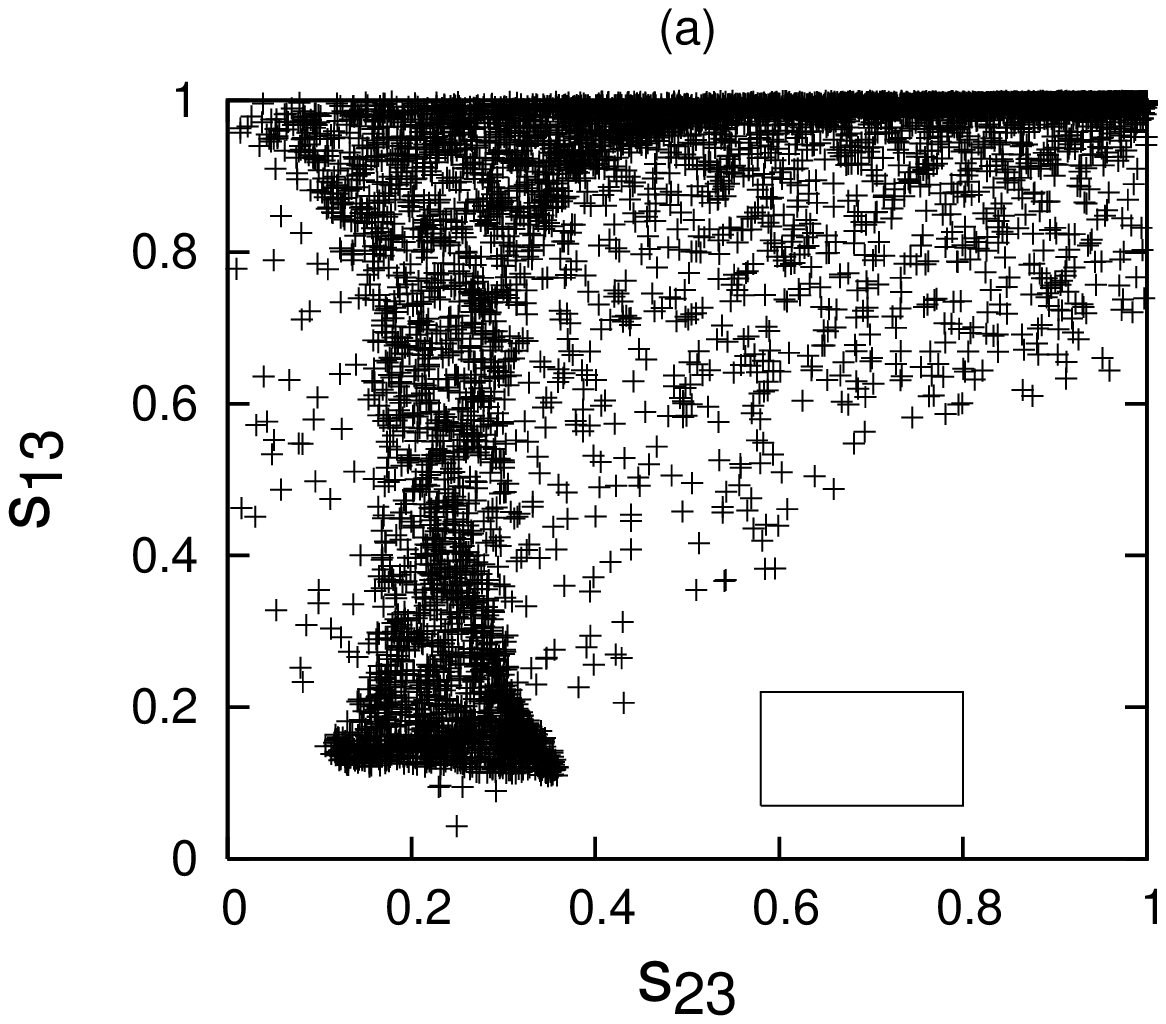}
    \end{minipage} \hspace{0.5cm}
\begin{minipage} {0.45\linewidth} \centering
\includegraphics[width=2.in,angle=-90]{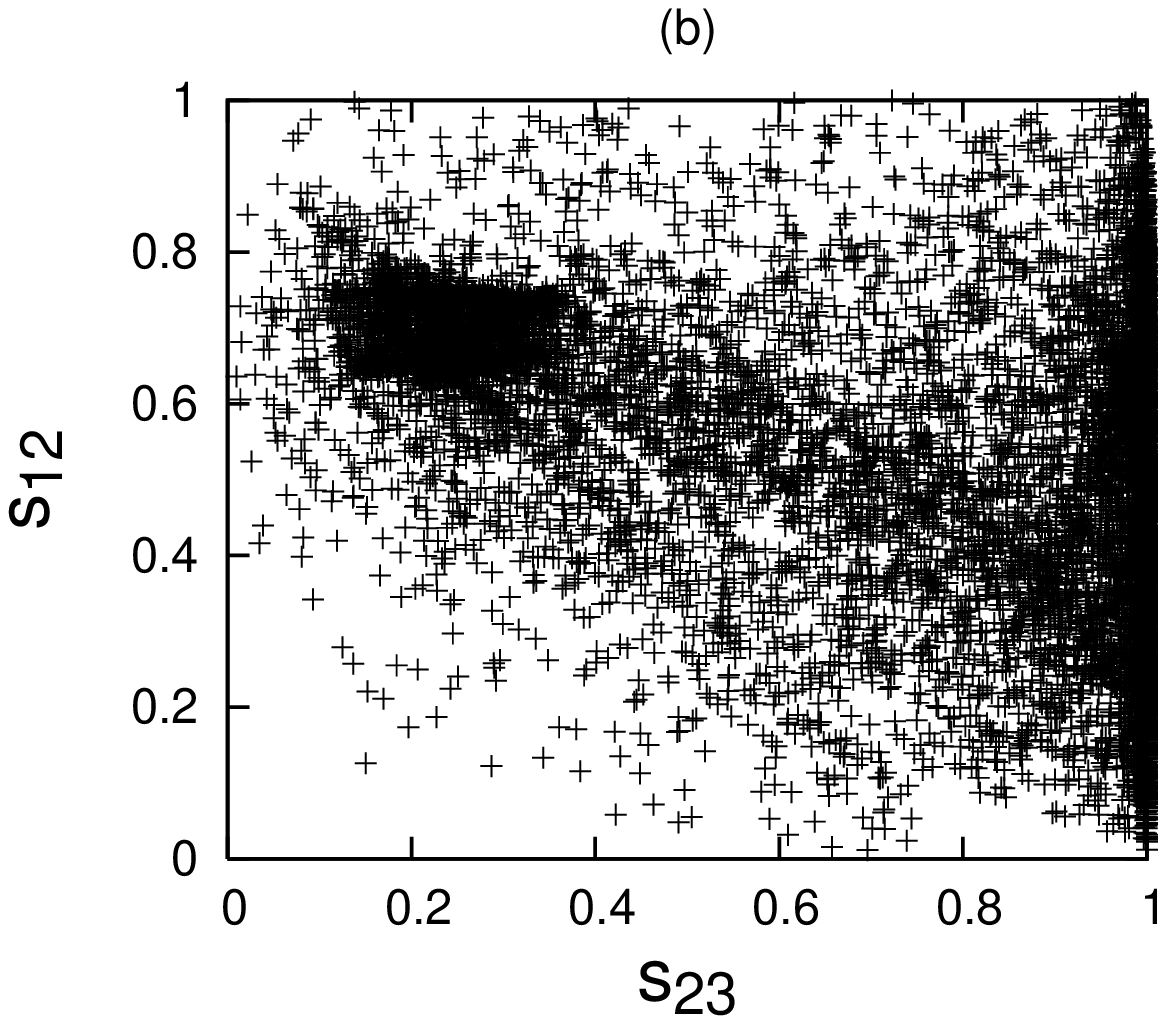}
  \end{minipage}\hspace{0.5cm}
  \begin{minipage} {0.45\linewidth} \centering
\includegraphics[width=2.in,angle=-90]{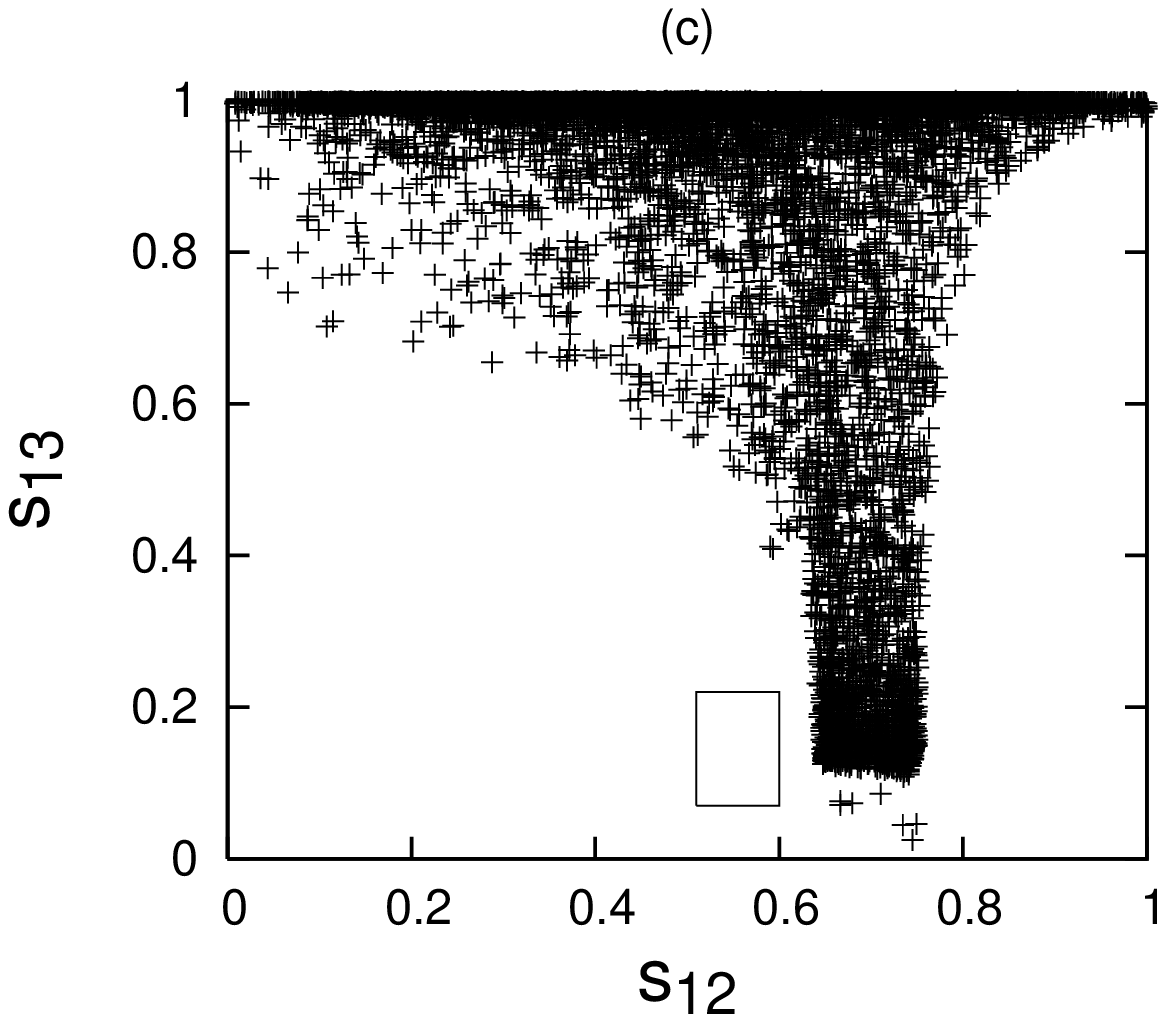}
  \end{minipage}\hspace{0.5cm}
   \caption{Plots showing the parameter space
corresponding to sines of any of the two mixing angles for texture
5 zero $D_l=0$ case of Dirac neutrino mass matrices}
\label{invdlz}
\end{figure}

Parallel to the case of texture 6 zero Dirac neutrino mass
matrices,  we would like to carry out similar analyses for the two
cases of texture 5 zero Dirac neutrino mass matrices as well. As a
first step, for the texture 5 zero $D_l=0$ case, in Figure
(\ref{invdlz}) we have plotted sines of any of the two mixing
angles for a particular value of $D_{\nu}= {m_{\nu_3}}$.

A general look at the graphs show that in the case of Figures
(\ref{invdlz}a) and (\ref{invdlz}c) the blank rectangular regions
show the experimentally allowed $3\sigma$ C.L. region of the
plotted angles, thus indicating towards ruling out of inverted
hierarchy of neutrino masses for this case. It may also be
mentioned that in Figure (\ref{invdlz}b), drawn for the sake of
completion, there seems to be an absence of a blank box thereby
indicating that for this case the experimentally allowed $3\sigma$
C.L. regions of the plotted angles overlap with each other,
therefore indicating towards viability of the inverted hierarchy
of neutrino masses. However, it may be noted that to rule out
inverted hierarchy, it is sufficient to do so from any of the
three plots of Figure (\ref{invdlz}).

\begin{figure}[hbt]
\begin{minipage}{0.45\linewidth}   \centering
\includegraphics[width=2.in,angle=-90]{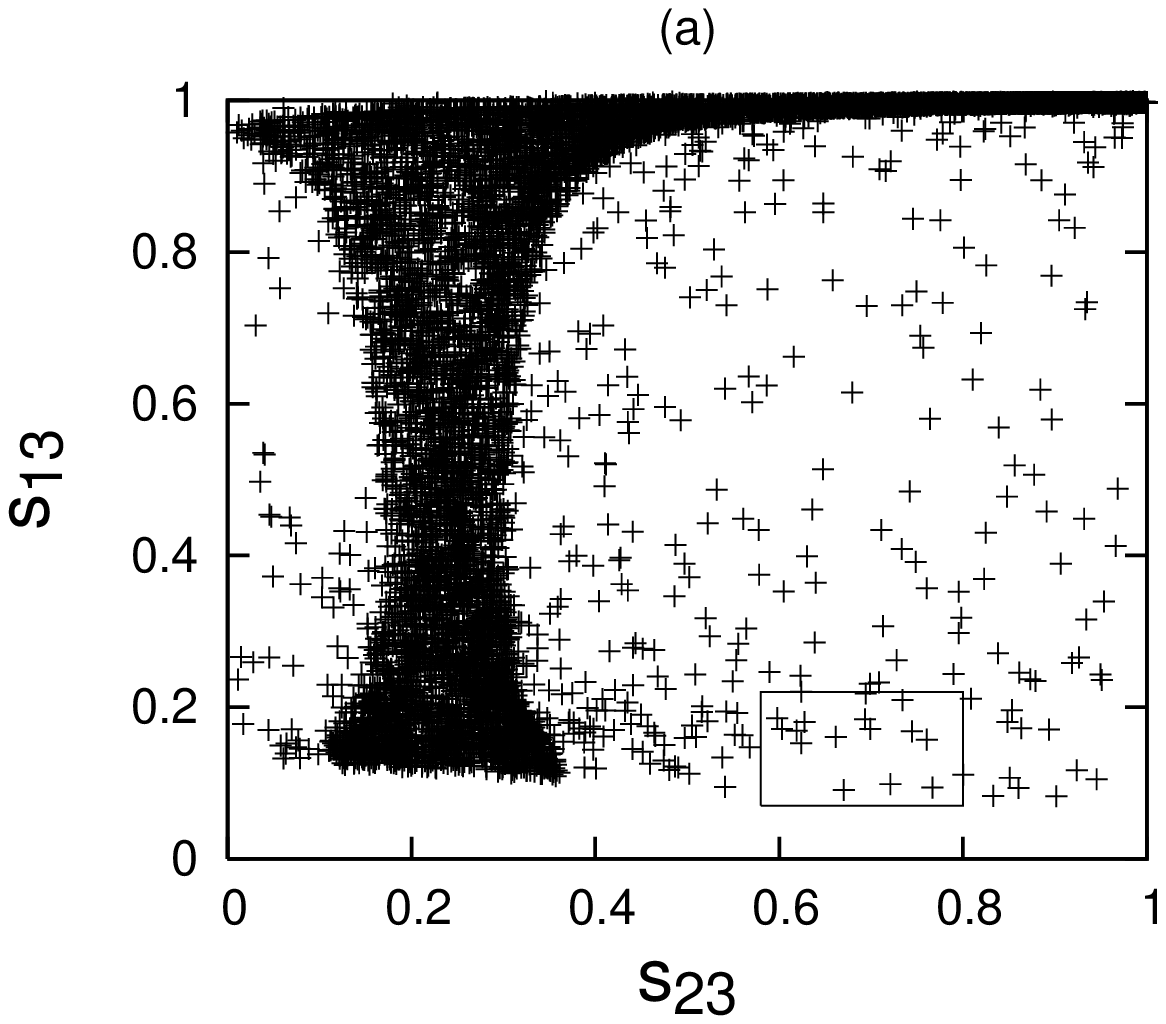}
    \end{minipage} \hspace{0.5cm}
\begin{minipage} {0.45\linewidth} \centering
\includegraphics[width=2.in,angle=-90]{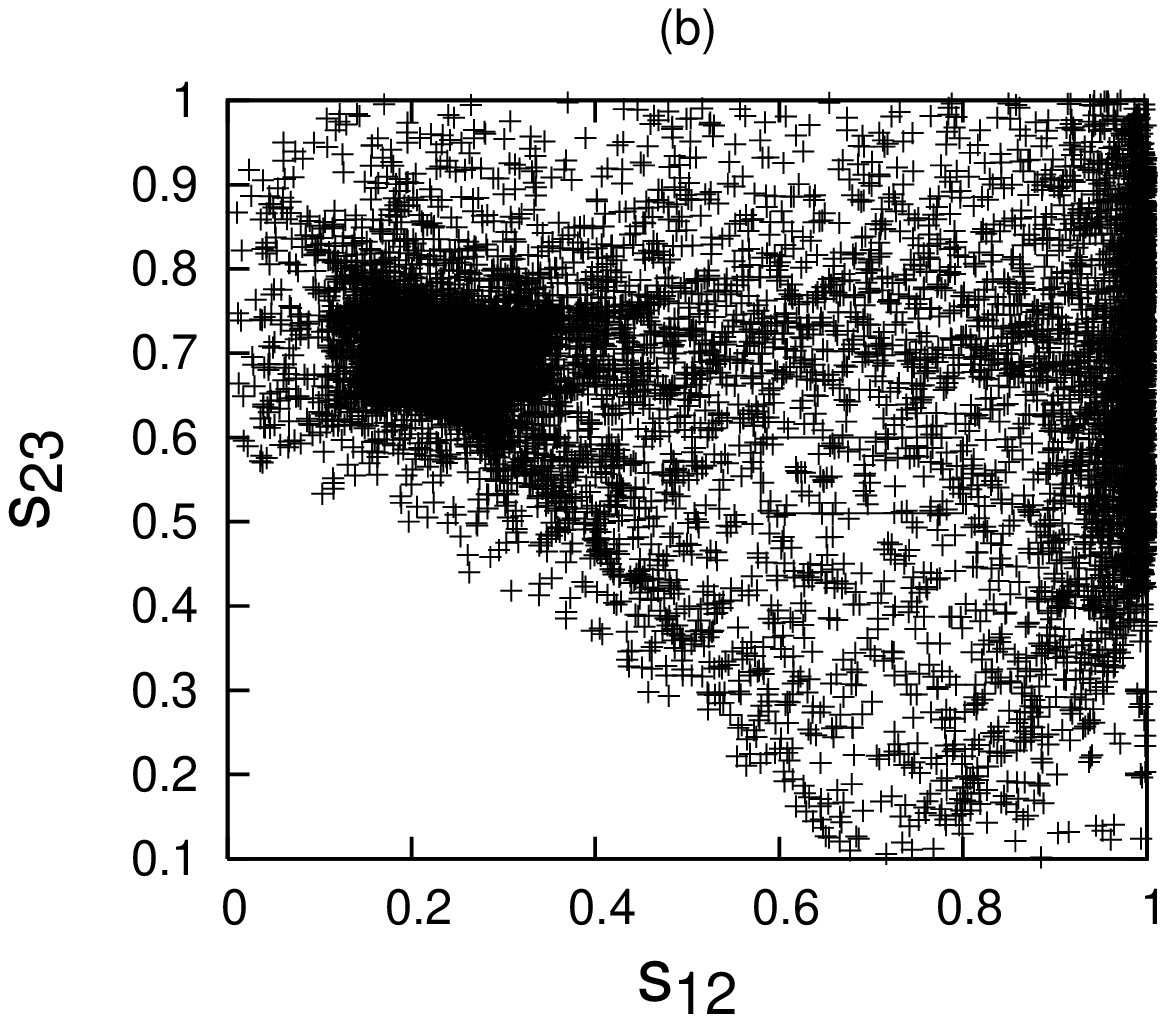}
  \end{minipage}\hspace{0.5cm}
  \begin{minipage} {0.45\linewidth} \centering
\includegraphics[width=2.in,angle=-90]{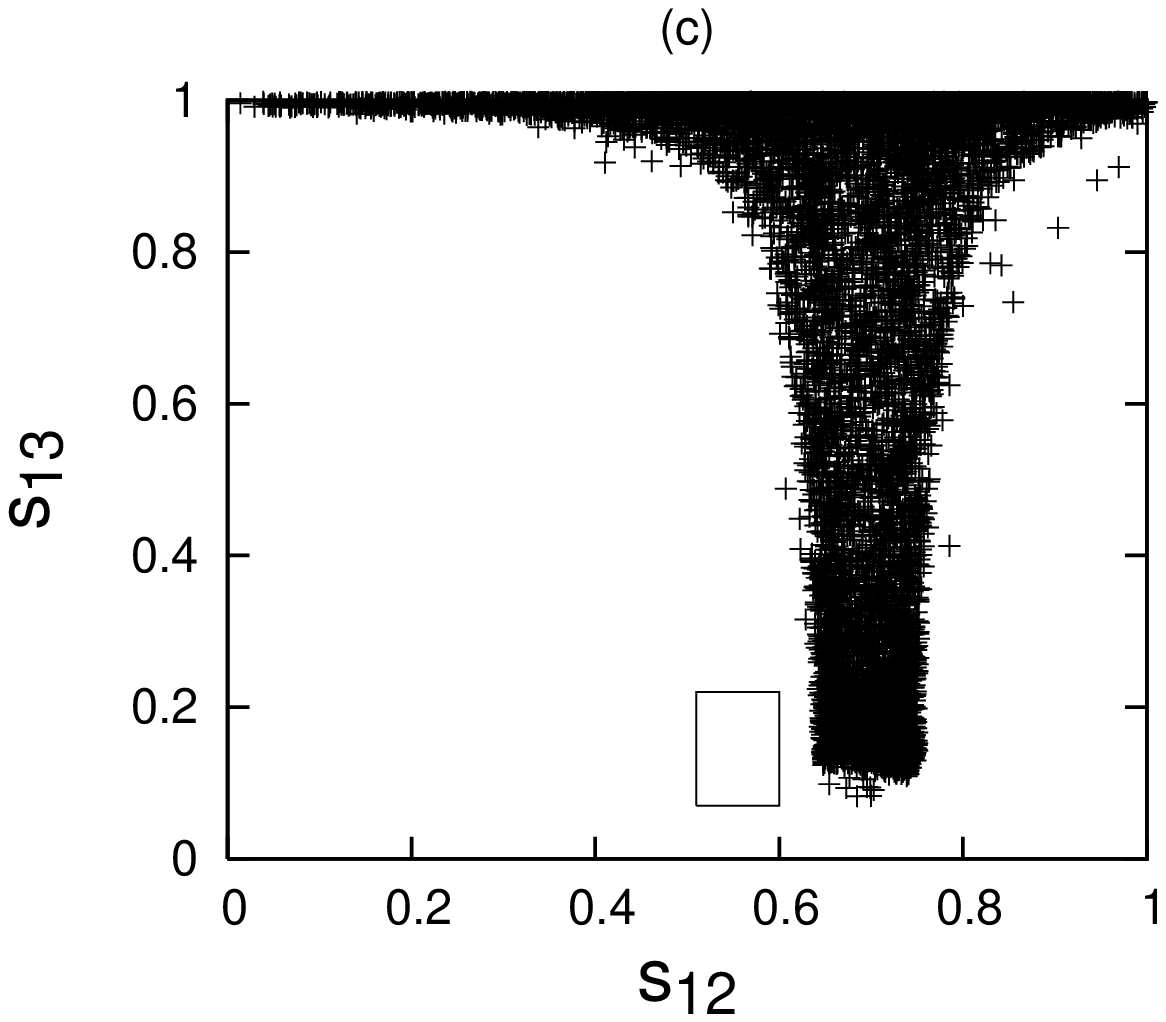}
  \end{minipage}\hspace{0.5cm}
 \caption{Plots showing the parameter space
corresponding to sines of any of the two mixing angles for texture
5 zero $D_{\nu}=0$ case of Dirac neutrino mass matrices}
\label{invdnz}
  \end{figure}

Coming to the next case of texture 5 zero Dirac neutrino mass
matrices, i.e., the $D_{\nu}=0$ case, again in Figure
(\ref{invdnz}) we have plotted sines of any of the two mixing
angles. Using Figure (\ref{invdnz}c) and similar arguments as
given for the previous cases, one can again conclude that this
case of texture 5 zero Dirac neutrino mass matrices is ruled out
for inverted hierarchy as well.

\subsection{Normal hierarchy of neutrino masses}

\begin{figure}[hbt]
\begin{minipage}{0.45\linewidth}   \centering
\includegraphics[width=2.in,angle=-90]{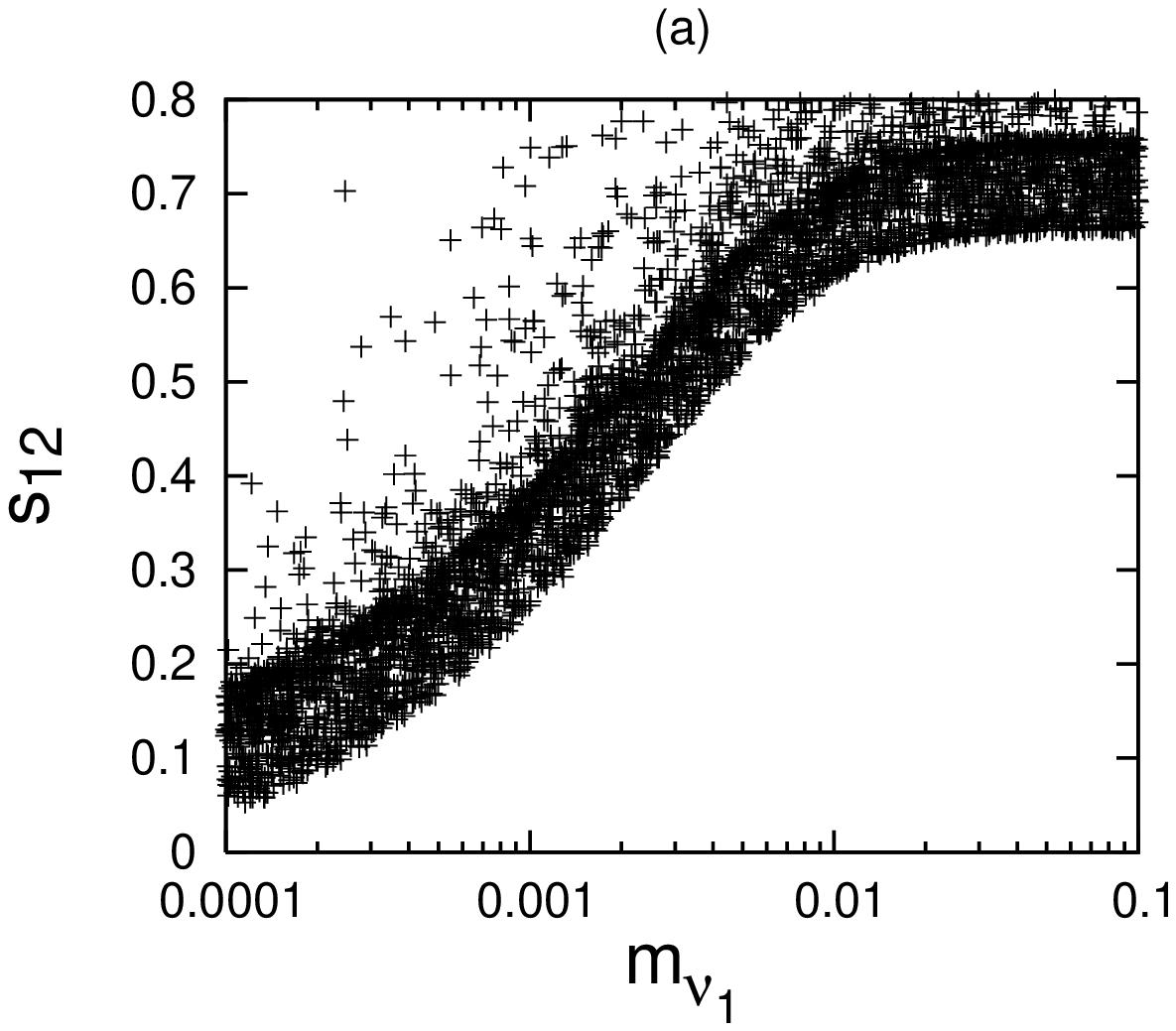}
    \end{minipage} \hspace{0.5cm}
\begin{minipage} {0.45\linewidth} \centering
\includegraphics[width=2.in,angle=-90]{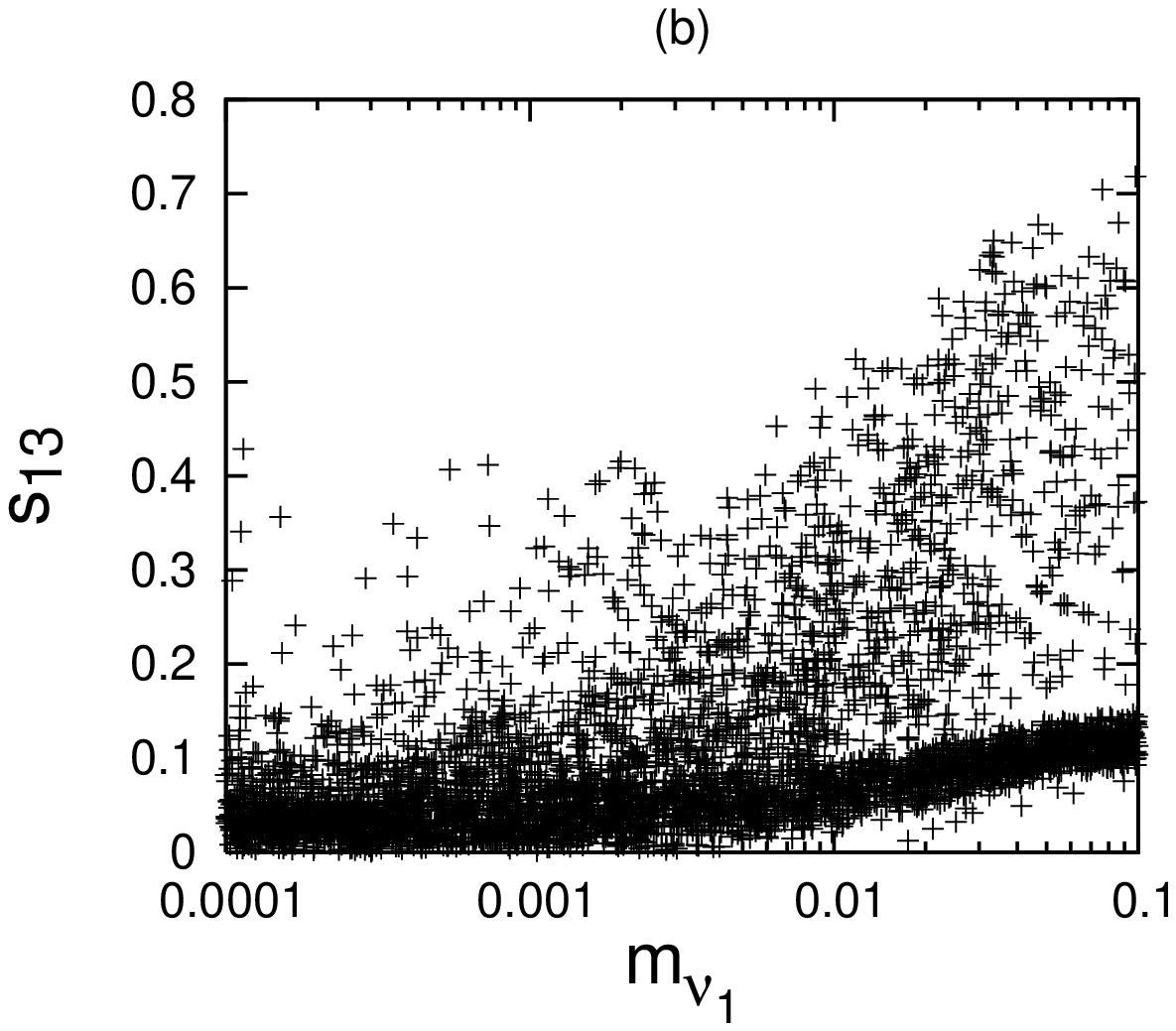}
  \end{minipage}\hspace{0.5cm}
  \begin{minipage}{0.45\linewidth}   \centering
\includegraphics[width=2.in,angle=-90]{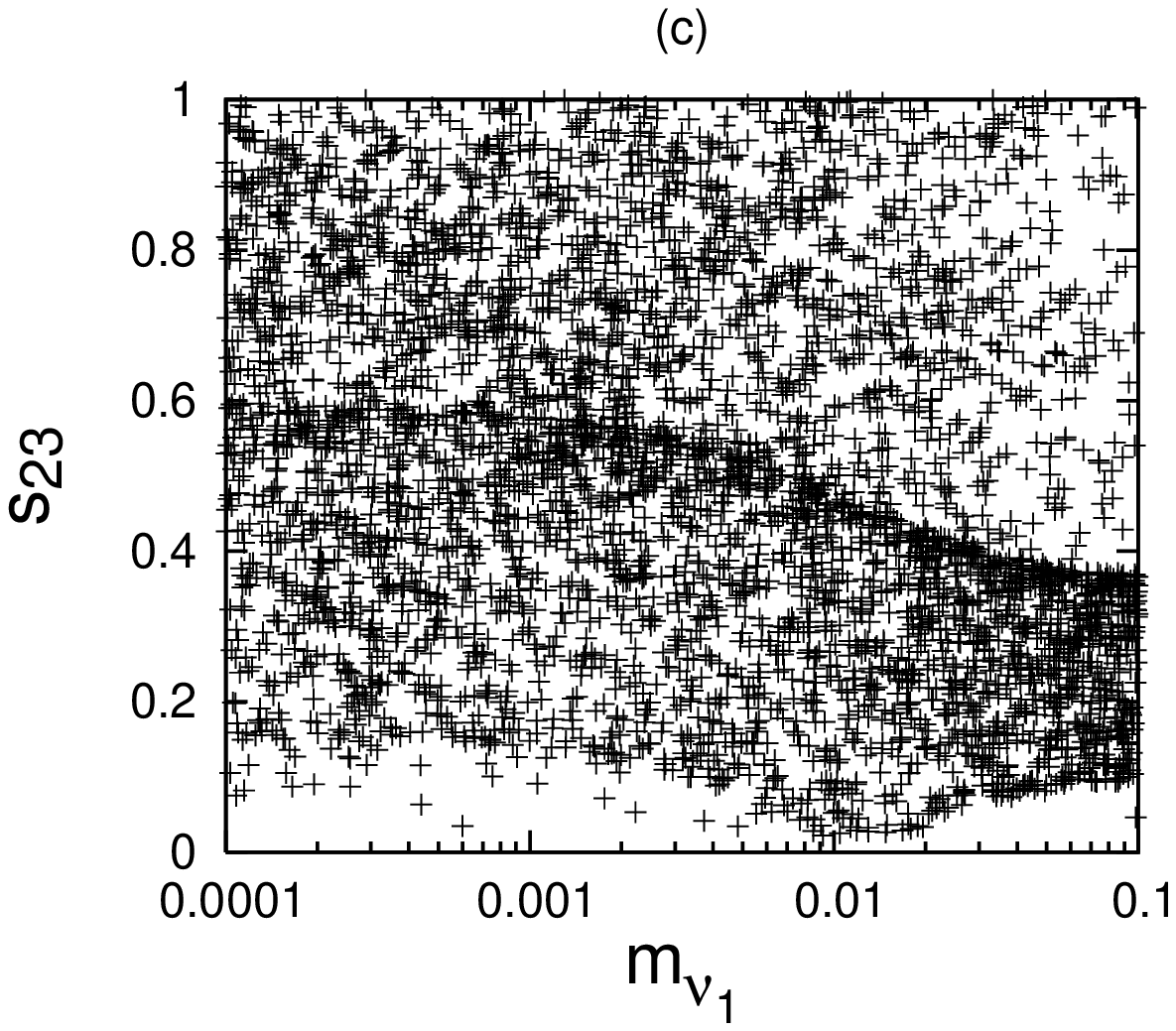}
    \end{minipage} \hspace{0.5cm}
\caption{Plots showing the variation of the three mixing angles
with the lightest neutrino mass $m_{\nu_1}$ for the $D_l= 0$ case
of texture 5 zero Dirac neutrino mass matrices}
  \label{nhdlz}
  \end{figure}

For the texture 5 zero mass matrices also we would like to explore
the implications of the three mixing angles on the neutrino mass
$m_{\nu_1}$. To this end, for the $D_l=0$ case, in Figure
(\ref{nhdlz}) we present the plots of the three mixing angles
versus $m_{\nu_1}$. The graphs shown in Figures (\ref{nhdlz}b) and
(\ref{nhdlz}c) do not seem to provide any constraints on the
neutrino mass $m_{\nu_1}$ due to the mixing angles $s_{13}$ and
$s_{23}$ respectively, however Figure (\ref{nhdlz}b) provides an
upper bound on $m_{\nu_1} \sim 0.01$eV.

For the sake of completion, we have also constructed the PMNS
matrix for the $D_{l}=0$ case of texture 5 zero mass matrices,
given as
 \be U=\left( \ba{ccc}
 0.765  -  0.869   &    0.489  -  0.623 &      0.006  -  0.198 \\
  0.182  -  0.467   &    0.518  -  0.728 &      0.584  -  0.809 \\
  0.310  -  0.579   &    0.421  -  0.684 &      0.573  -   0.806

 \ea \right). \label{dirmat}\ee
Interestingly, the above matrix shows good deal of compatibility
with a recently constructed PMNS matrix by Garcia \cite{garcia2}.
\begin{figure}[hbt]
\begin{minipage}{0.45\linewidth}   \centering
\includegraphics[width=2.in,angle=-90]{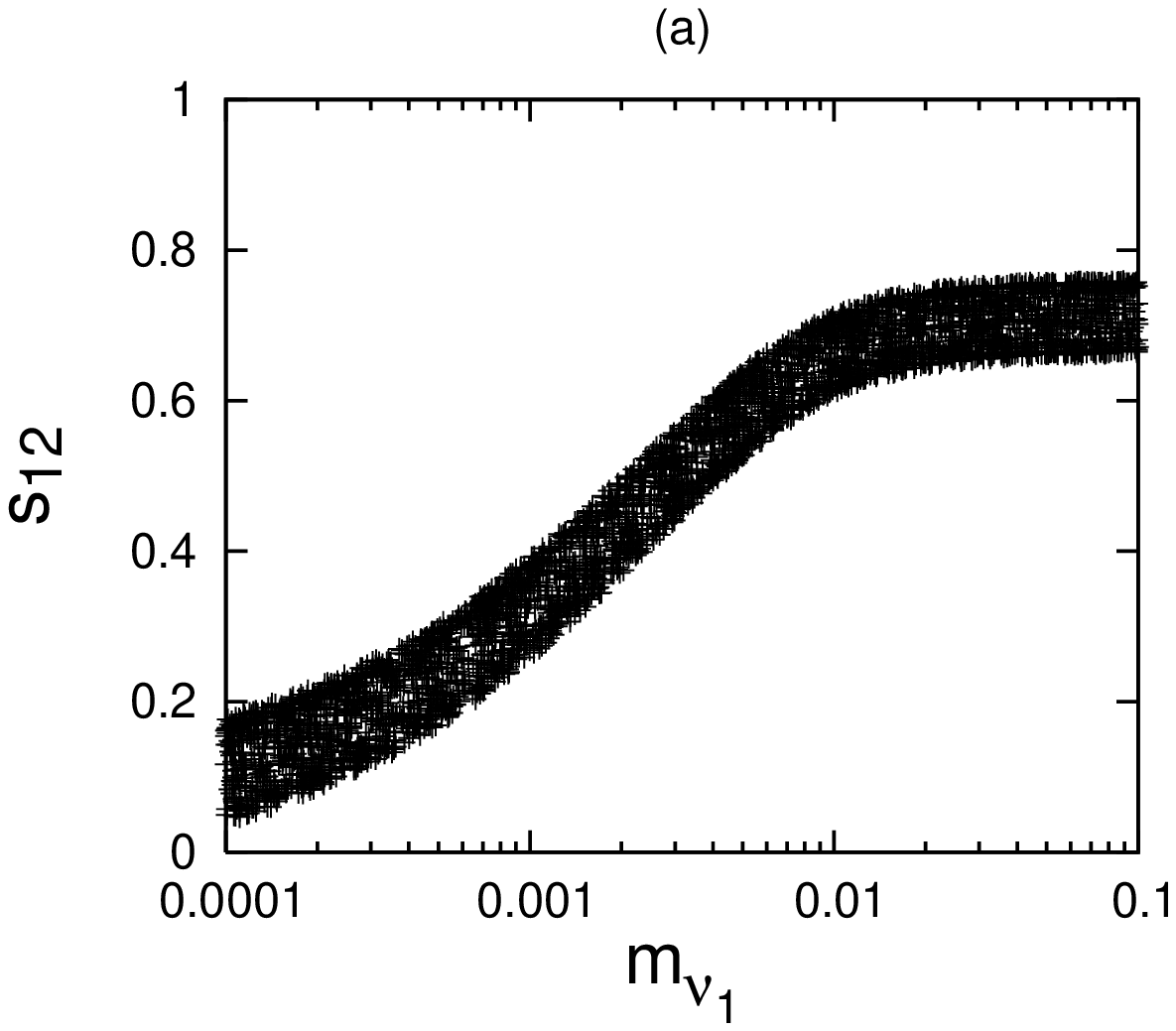}
    \end{minipage} \hspace{0.5cm}
\begin{minipage} {0.45\linewidth} \centering
\includegraphics[width=2.in,angle=-90]{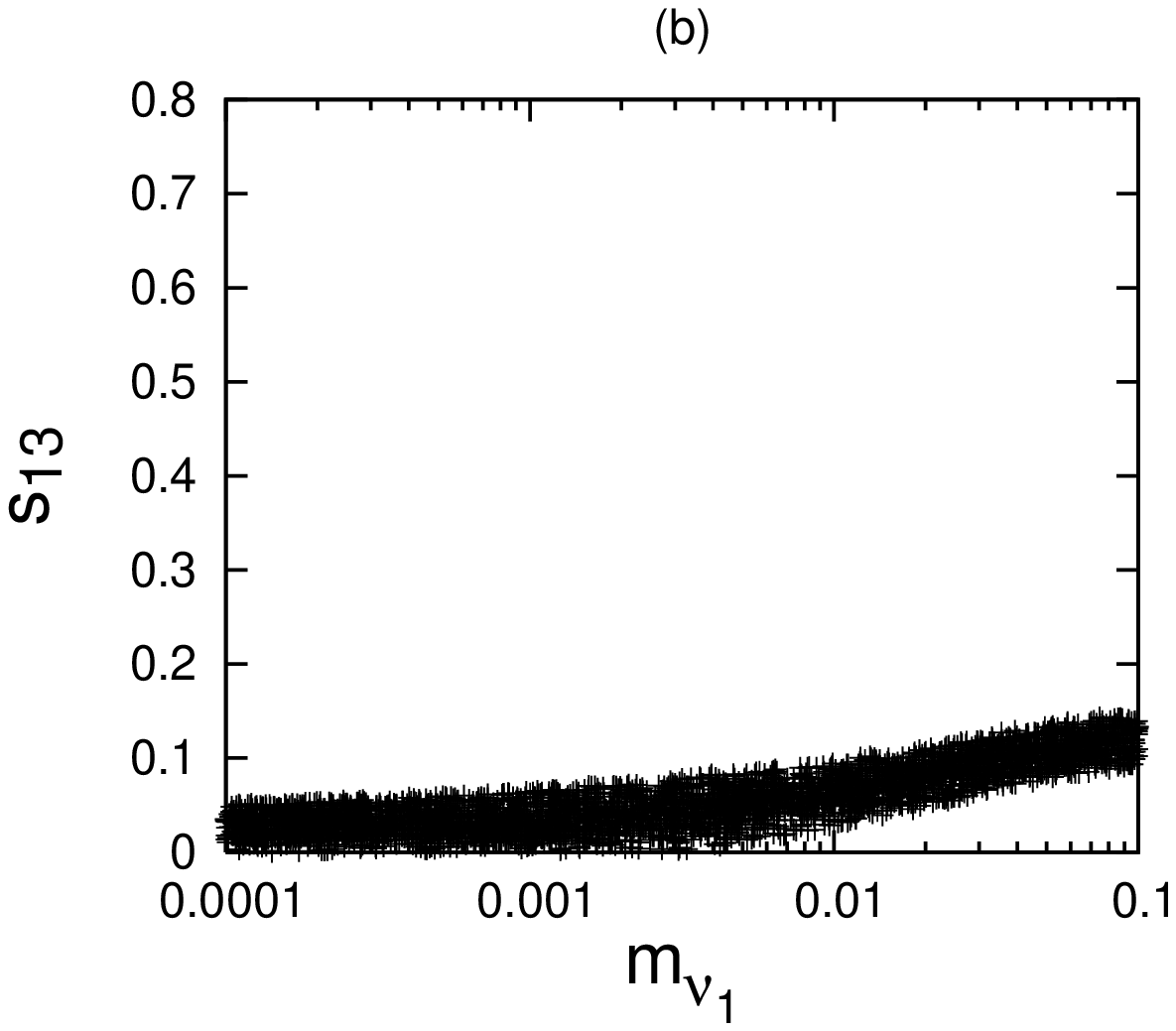}
  \end{minipage}\hspace{0.5cm}
  \begin{minipage}{0.45\linewidth}   \centering
\includegraphics[width=2.in,angle=-90]{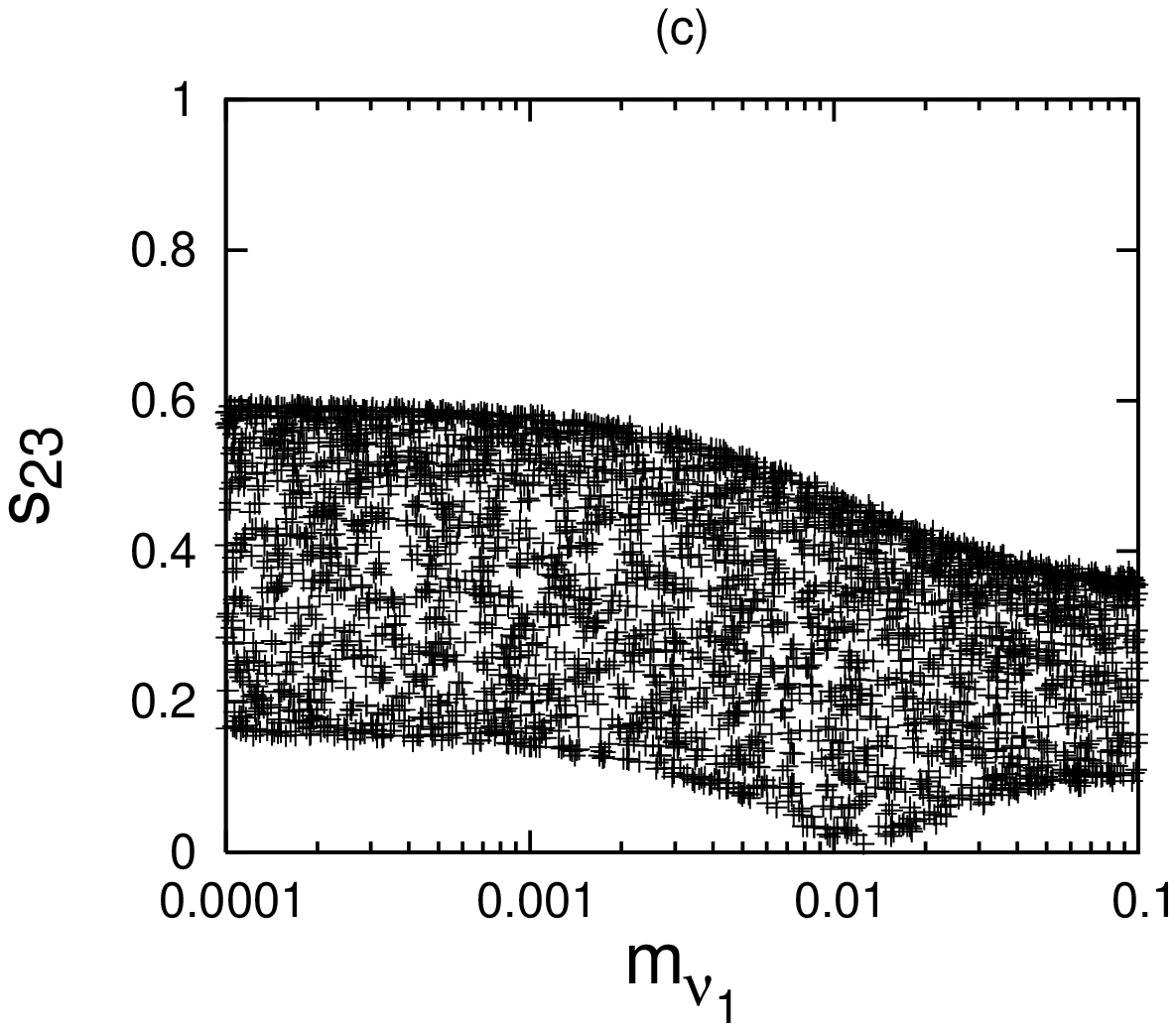}
    \end{minipage} \hspace{0.5cm}
\caption{Plots showing the variation of the three mixing angles
with the lightest neutrino mass $m_{\nu_1}$ for the $D_{\nu}= 0$
case of texture 5 zero Dirac neutrino mass matrices}
  \label{nhdnz}
  \end{figure}Similarly, for the $D_{\nu}=0$ case of texture 5 zero Dirac neutrino
mass matrices, in Figure (\ref{nhdnz}) we have plotted the graphs
showing the variation of the three neutrino mixing angles w.r.t.
$m_{\nu_1}$. Interestingly, from a general look at the plots one
finds that the $s_{13}$ and $s_{23}$ versus $m_{\nu_1}$ graphs,
Figures (\ref{nhdnz}b) and (\ref{nhdnz}c), are very similar to the
corresponding plots of texture 6 zero case. It may be noted that
the normal hierarchy of neutrino masses for texture 6 zero Dirac
neutrino mass matrices has already been ruled out, on similar
lines Figures (\ref{nhdnz}b) and (\ref{nhdnz}c) corresponding to
the $1\sigma$ C.L. range of $s_{13}$, indicate towards the ruling
out of normal hierarchy for Dirac neutrinos.

\begin{figure}[hbt]
\begin{minipage}{0.45\linewidth}   \centering
\includegraphics[width=2.in,angle=-90]{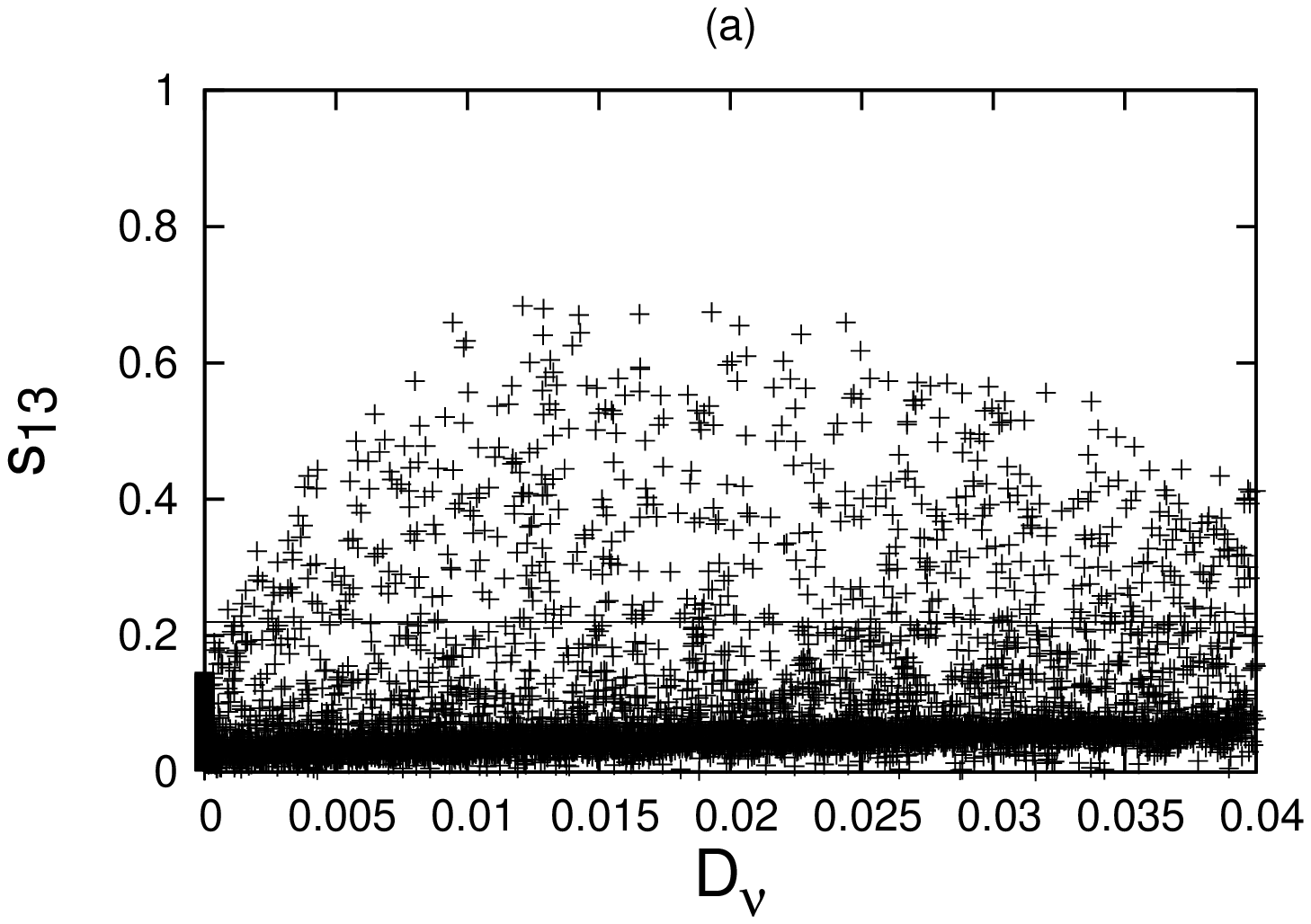}
    \end{minipage} \hspace{0.5cm}
\begin{minipage} {0.45\linewidth} \centering
\includegraphics[width=2.in,angle=-90]{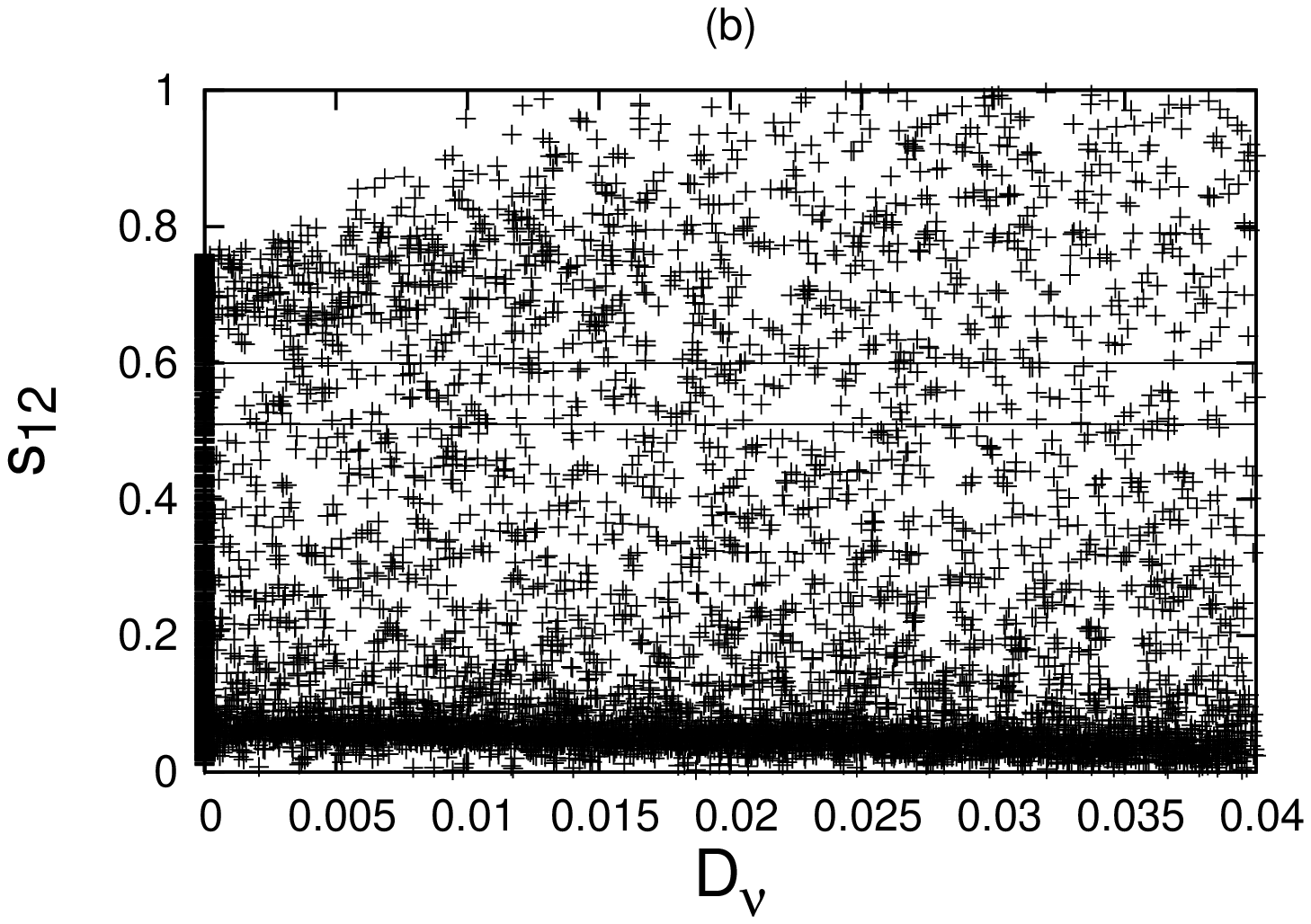}
  \end{minipage}\hspace{0.5cm}
  \begin{minipage}{0.45\linewidth}   \centering
\includegraphics[width=2.in,angle=-90]{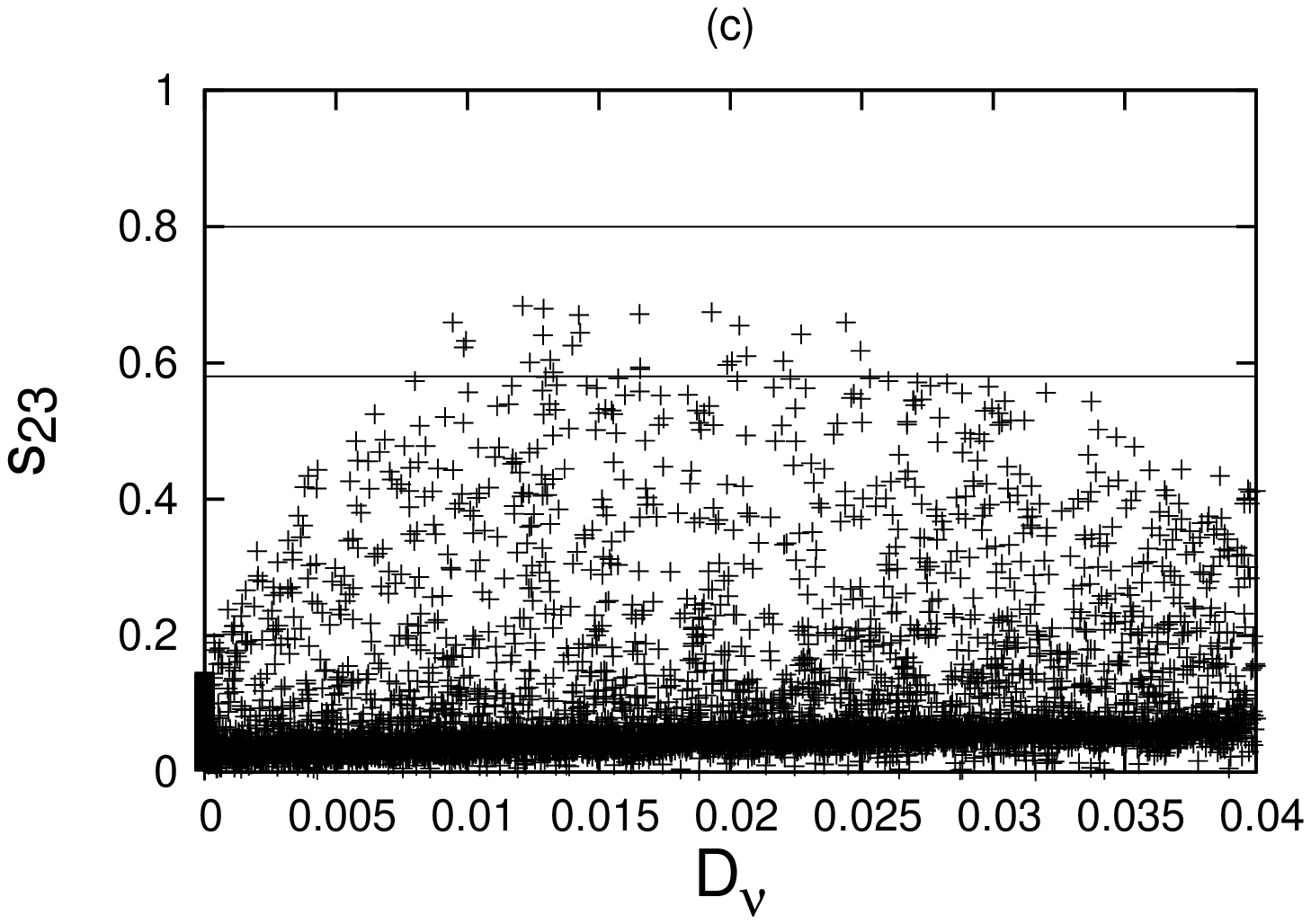}
    \end{minipage} \hspace{0.5cm}
   \caption{Plots
showing variation of $ s_{13} $, $ s_{12} $ and $ s_{23} $ with
the for texture 5 zero Dirac neutrinos for the $ D_{l}= 0 $ case
for normal hierarchy.}
  \label{dnuplots}
  \end{figure}

Comparing the two cases of texture 5 zero Dirac neutrino mass
matrices, one finds that out of the two free parameters of the
mass matrices $D_l$ and $D_{\nu}$, the parameter $D_{\nu}$ plays a
more important role in establishing the compatibility of texture 5
zero Dirac neutrino mass matrices. The variation of this parameter
with the three mixing angles has been examined and these plots
have been presented in Figure (\ref{dnuplots}). A general look at
these plots reveals that the mixing angles $s_{13}$ and $s_{12}$
seem to hardly put any restrictions on the possible values of
$D_{\nu}$. However, the angle $s_{23}$ provides a constraint on
the $D_{\nu}$ values, for example, from Figure (\ref{dnuplots}c)
one finds $D_{\nu} \sim 0.01-0.03$eV.

\subsection{Degenerate scenario of neutrino masses}
Coming to the case of degenerate scenario of neutrino masses for
the two cases of texture 5 zero mass matrices. Parallel to the
degenerate scenario for texture 6 zero mass matrices, for the
$D_{\nu}=0$ case also the degenerate scenario corresponding to
both normal and inverted hierarchy seems to be ruled out. This can
be understood  by noting that since both normal and inverted
hierarchy of neutrino masses are already ruled out, therefore, the
corresponding degenerate scenarios are also ruled out.

For the $D_{l}=0$ case, again the degenerate scenario
corresponding to inverted hierarchy is ruled out since for this
case inverted hierarchy of neutrino masses is already ruled out.
For the degenerate scenario following normal hierarchy of neutrino
masses which is viable for this case of texture 5 zero mass
matrices, again Figure (\ref{nhdnz}a) can be used to rule it out.
From the figure one finds that for $m_{\nu_1}$ around
$0.1~\rm{eV}$, there is no overlap of the plotted angle $s_{12}$
with its experimental limits.

\section{Summary and conclusions}
To summarize, we have carried out detailed calculations pertaining
to three cases, i.e., two possible cases of texture 5 zero
Fritzsch-like hermitian lepton mass matrices, $D_l=0$ case and
$D_{\nu}=0$ case. Corresponding to each of these case, we have
considered three possibilities of neutrino masses having
normal/inverted hierarchy and degenerate scenario. The detailed
dependence of mixing angles on the lightest neutrino mass have
been investigated for texture 6 zero as well as for texture 5 zero
cases.

The analysis leads to several interesting results. For Dirac
neutrinos, all the cases pertaining to inverted hierarchy and
degenerate scenario of neutrino masses have been ruled out for
texture 5 zero  mass matrices. Interestingly, in the case of
 texture 5 zero $D_{\nu}=0$
case, the normal hierarchy of neutrino masses is also ruled out at
$1\sigma$ C.L.. Refinements in the data can make these conclusion
more rigorous.

Corresponding to the texture 5 zero $D_{l}=0$ case, the normal
hierarchy of neutrino masses is viable and the plot of the mixing
angle $s_{12}$ versus $m_{\nu_1}$ provides an upper bound on
$m_{\nu_1} \sim 0.01$eV. The PMNS matrix for this case has also
been constructed which shows good deal of compatibility with a
recently constructed PMNS matrix by Garcia \cite{garcia2}.
Further, one finds that out of the two free parameters of the mass
matrices, $D_l$ and $D_{\nu}$, the parameter $D_{\nu}$ plays a
more important role in establishing the compatibility of texture 5
zero Dirac neutrino mass matrices. In this context, variation of
$D_{\nu}$ with the three mixing angles has been examined and one
finds that the angle $s_{23}$ provides a constraint on $D_{\nu}
\sim 0.01-0.03$eV.

\textbf{Acknowledgements} \vskip0.05 cm P.F. would like to
acknowledge Principal D.A.V. College and S.S. would like to
acknowledge Principal GGDSD College along with Chairperson,
Department of Physics, P.U. for providing facilities to work. The
authors would also like to acknowledge M. Gupta for useful
discussions.

\end{document}